\documentclass[twocolumn,showpacs,preprintnumbers,amsmath,amssymb]{revtex4}

\usepackage{epsfig}
\usepackage{graphicx}
\usepackage{dcolumn}
\usepackage{bm}

\let\jnfont=\rm
\def\NPB#1,{{\jnfont Nucl.\ Phys.\ B }{\bf #1},}
\def\PLB#1,{{\jnfont Phys.\ Lett.\ B }{\bf #1},}
\def\EPJC#1,{{\jnfont Eur.\ Phys.\ Jour.\ C }{\bf #1},}
\def\PRD#1,{{\jnfont Phys.\ Rev.\ D }{\bf #1},}
\def\PRL#1,{{\jnfont Phys.\ Rev.\ Lett.\ }{\bf #1},}
\def\MPLA#1,{{\jnfont Mod.\ Phys.\ Lett.\ A }{\bf #1},}
\def\JPG#1,{{\jnfont J.\ Phys.\ G}{\bf #1},}
\def\CTP#1,{{\jnfont Commun.\ Theor.\ Phys.\ }{\bf #1},}

\begin{document}

\preprint{hep-ph/0612273}

\title{Virtual Effects of Split SUSY in Higgs Productions at Linear Colliders}

\author{Fei Wang$^1$, Wenyu Wang$^2$, Fuqiang Xu$^2$, Jin Min Yang$^{3,2}$, Huanjun Zhang$^{2,4}$}

\affiliation{
   $^1$ Center for High Energy Physics, Tsinghua University, Beijing 100084, China\\
   $^2$ Institute of Theoretical Physics, Academia Sinica, Beijing 100080, China\\
   $^3$ CCAST (World Laboratory), P.O.Box 8730, Beijing 100080, China \\
   $^4$ Department of Physics, Henan Normal University, Xinxiang 453007,  China }

\date{\today}

\begin{abstract}
In split supersymmetry the gauginos and higgsinos are the only supersymmetric
particles possibly accessible at foreseeable colliders like the CERN Large Hadron Collider
(LHC) and the International Linear Collider (ILC).
In order to account for the cosmic dark matter measured by WMAP, these gauginos and higgsinos
are stringently constrained and could be explored at the colliders through
their direct productions and/or virtual effects in some processes.
The clean environment and high luminosity of the ILC render the virtual effects
of percent level meaningful in unraveling the new physics effects.
In this work we assume split supersymmetry and calculate the virtual effects
of the WMAP-allowed gauginos and higgsinos in Higgs productions $e^+e^-\to Z h$ and
$e^+e^-\to \nu_e \bar\nu_e h$  through $WW$ fusion at the ILC.
We find that the production cross section of $e^+e^-\to Zh$ can be altered by a
few percent in some part of the WMAP-allowed parameter space, while
the correction to the $WW$ fusion process $e^+e^-\to \nu_e \bar\nu_e h$
is below $1\%$. Such virtual effects are correlated with
the cross sections of chargino pair productions and can offer complementary
information in probing split supersymmetry at the colliders.
\end{abstract}


\maketitle

\section{Introduction}
Since supersymmetry (SUSY) is so appealing in particle physics, cosmology and
string theory, its exploration will be a central focus of future collider
experiments. If SUSY is at TeV-scale, as required by solving the fine-tuning
problem in particle physics, the LHC expects to discover it or at least reveal some
of its fingerprints and then the ILC \cite{ILC} will zero in on its precision test and map out its
detailed structure. However, if the fine-tuning in particle physics works in nature,
just like the fine-tuning for the cosmological constant, SUSY may turn out to be
a kind of split-SUSY \cite{split}, in which all scalar supersymmetric
particles (sfermions and additional Higgs bosons) are superheavy and only
gauginos and higgsinos are possibly light and accessible at foreseeable colliders
like the LHC and ILC.  So, if split-SUSY is the true story, the focus of
experimental and theoretical studies on SUSY will be gauginos and higgsinos.

To facilitate the collider searches for gauginos and higgsinos in split-SUSY,
it is important to examine the possible range of their masses by considering
various direct and indirect constraints and requirements. The lightness of
gauginos and higgsinos is required by the consideration of the unification
of gauge couplings and the explanation of cosmic dark matter. It turns out
that the gauge coupling unification does not require gauginos or higgsinos
necessarily below TeV scale and they may be as heavy as 10 TeV \cite{senatore,wang}.
However, the cosmic dark matter measured by WMAP imposes much stronger
constraints on the masses of gauginos and higgsinos (except gluinos),
whose lightest mass eigenstates, i.e., the lightest neutralino and chargino,
must be lighter than about 1 TeV under the popular assumption $M_1=M_2/2$
with $M_1$ and $M_2$ being the U(1) and SU(2) gaugino masses, respectively
\cite{wenyu,pierce,profumo}.

Note that unlike the neutralinos and charginos, the gluino is not directly subject to
the dark matter constraints and its mass constrained by gauge coupling unification
can be as high as 18 TeV \cite{senatore}. Theoretically, the gluino is usually speculated
to be much heavier than neutralinos and charginos.  So, although the gluino is the only
colored particle among gauginos and higgsinos and usually expected to be copiously produced
in the gluon-rich environment of the LHC \cite{gluino-lhc},
it may be quite heavy and thus out of the reach of
the LHC and ILC. Therefore, to explore split-SUSY, it is important to examine
the neutralinos and charginos.

The neutralinos and charginos in split-SUSY constrained by the cosmic dark matter
can be explored at the LHC and ILC in two ways. One way is directly looking for their
productions, such as chargino pair productions.
Our previous analysis \cite{wenyu} showed
that the chargino pair production rates at the LHC and ILC are quite large in some part
of the WMAP-allowed parameter space, but in the remained part of the parameter space the
production rates are unobservably small. The other way to reveal the existence of these
particles is through disentangling their virtual effects in some processes
which can be precisely measured. It is shown that SUSY may have
sizable virtual effects in Higgs boson processes  \cite{higgs-residue}
and top quark processes \cite{top-susy} since they are the heaviest particles in
the SM and sensitive to new physics.
For split-SUSY, its virtual effects in top quark interactions and
Higgs-fermion Yukawa interactions are expected to be small
since the relevant vertex loops always involve sfermions which
are superheavy.
So, to reveal the virtual effects of split-SUSY, we concentrate on the gauge interactions
of the Higgs boson.  Such virtual effects of weakly interacting
neutralinos and charginos are usually at percent level and only the high-luminosity $e^+e^-$
collider like the ILC can possibly have such percent-level sensitivity.
As the discovery machine, the LHC, however, is not expected to be able to disentangle
such percent-level quantum effects due to its messy hadron backgrounds.
So in this work we investigate the virtual effects
of the WMAP-allowed split-SUSY in Higgs productions $e^+e^-\to Z h$ and
$e^+e^-\to \nu_e \bar\nu_e h$  through $WW$ fusion at the ILC.
Note that although the SUSY corrections to these processes were calculated in
the literature \cite{complete-susy,rewwh}, our studies in this work are still necessary
since those calculations were performed in the framework of the general minimal
supersymmetric model and did not consider the dark matter constraints.

This work is organized in the follows.
In Sec. II we calculate the split-SUSY loop contributions to Higgs production
$e^+e^-\to Z h$ and $e^+e^-\to \nu_e \bar\nu_e h$  through $WW$ fusion at the ILC.
In Sec. III we present some numerical results for the parameter space under
WMAP dark matter constraints. The conclusion is given in Sec. IV.
Note that for the SUSY parameters we adopt the notations in \cite{gunion}.
We assume the lightest supersymmetric particle is the lightest neutralino,
which solely makes up the cosmic dark matter.

\section{Calculations}
\subsection{About split-SUSY}
In split-SUSY the Higgs sector at low energy is fine-tuned to have only
one Higgs doublet \cite{split} and the effective spectrum of superparticles
contains the higgsinos $\tilde H_{u,d}$, winos $\tilde W^i$,  bino $\tilde B$
and gluino $\tilde{g}$. The most genenral renormalizable Lagrangian at low
energy (say TeV scale) contains the interactions
\begin{eqnarray}\label{lagran}
{\mathcal L}&=&m^2 H^\dagger H-\frac{\lambda}{2}
   \left( H^\dagger H\right)^2  \nonumber \\
&& -\left[ h^u_{ij} \bar{q}_j u_i\epsilon H^* +h^d_{ij} \bar{q}_j d_iH
         +h^e_{ij} \bar{\ell}_j e_iH \right. \nonumber \\
&& +\frac{M_3}{2} \tilde{g}^A \tilde{g}^A
   +\frac{M_2}{2} \tilde {W}^a \tilde{W}^a
   +\frac{M_1}{2} \tilde{B}\tilde{B} \nonumber \\
&&  \left.  +\mu \tilde{H}_u^T\epsilon\tilde{H}_d
   +\frac{H^\dagger}{\sqrt{2}}\left( \tilde{g}_u \sigma^a {\tilde W}^a
   +\tilde{g}^\prime_u\tilde{B} \right)\tilde{H}_u \right. \nonumber \\
&&   \left.+\frac{H^T\epsilon}{\sqrt{2}}\left(
   -\tilde{g}_d \sigma^a\tilde{W}^a+\tilde{g}^\prime_d\tilde{B} \right) \tilde{H}_d
   +\rm {h.c.}\right] ,
\end{eqnarray}
where $\epsilon =i\sigma_2$. Thus the Higgs sector in split-SUSY is same as
in the SM except for the additional Higgs couplings to gauginos and higgsinos.
Other four Higgs bosons in the MSSM are superheavy and decouple.
As is well known,
an upper bound of about 135 GeV exists for the lightest Higgs boson in the
MSSM \cite{higgsmass}, which is relaxed to about 150 GeV in split-SUSY \cite{split}.

The gauginos (winos and bino) and higgsinos mix into the mass eigenstates called charginos and neutralinos.
The chargino mass matrix is given by
\small
\begin{eqnarray} \label{mass1}
\left( \begin{array}{cc} M_2 & \sqrt 2 m_W \sin\beta \\
                        \sqrt 2 m_W \cos\beta & \mu \end{array} \right) ,
\end{eqnarray}
and the neutralino mass matrix is given by
\begin{eqnarray} \label{mass2}
\left( \begin{array}{cccc}
M_1 & 0   & - m_Z s_W c_\beta & m_Z s_W s_\beta  \\
0   & M_2 & m_Z c_W c_\beta   &-m_Z c_Ws_\beta  \\
- m_Z s_W c_\beta &  m_Z c_W c_\beta  & 0 & -\mu \\
m_Z s_W s_\beta & -m_Z c_W s_\beta & -\mu & 0\\
\end{array} \right),
\end{eqnarray}
\normalsize
where $s_W=\sin\theta_W$ and $c_W=\cos\theta_W$ with $\theta_W$ being the weak mixing angle,
and $s_\beta=\sin\beta$ and $c_\beta=\cos\beta$ with $\beta$ defined by
$\tan\beta = v_2/v_1$, the ratio of the vacuum expectation values
of the two Higgs doublets.
$M_1$ and $M_2$ are respectively the $U(1)$ and $SU(2)$ gaugino mass parameters,
and $\mu$ is the mass parameter in the mixing term $-\mu \epsilon_{ij} H_u^iH_d^j$ in the
superpotential. The diagonalization of (\ref{mass1}) gives two
charginos $\tilde \chi^+_{1,2}$ with the convention $M_{\tilde\chi^+_1}<M_{\tilde\chi^+_2}$;
while the diagonalization of (\ref{mass2}) gives four neutralinos $\tilde\chi^0_{1,2,3,4}$
with the convention $M_{\tilde\chi^0_1}<M_{\tilde\chi^0_2}<M_{\tilde\chi^0_3}<M_{\tilde\chi^0_4}$.
So the masses and mixings of charginos and neutralinos
are determined by four parameters: $M_1$, $M_2$, $\mu$ and $\tan\beta$.

Note that the low energy lagrangian in Eq.(\ref{lagran})
should be understood as an effective theory after squarks, sleptons, and
heavier Higgs bosons are integrated out.
Then, as is discussed in \cite{split}, the Higgs-higgsino-gaugino
couplings in Eq.(\ref{lagran}) should deviate from the SUSY results shown
in the off-diagonal elements of the mass matrices in Eqs.(2) and (3),
although such deviation is negligible for numerical results.

In split SUSY the possible channels of Higgs ($h$) productions at
the ILC are the Higgs-strahlung process $e^+e^-\to Z^* \to Z h$
and $WW$-fusion process $e^+e^-\to \nu_e \bar\nu_e h$. Both
processes will be precisely measured at the ILC if the light Higgs
boson $h$ is indeed found at the LHC. Since these processes may be
sensitive to new physics, they may serve as a good probe for
TeV-scale new physics. Other channels, such as the production of
$h$ associated with a CP-odd Higgs boson $A$ and the charged Higgs
pair production, cannot occur due to the superheavy $A$ and the
superheavy charged Higgs bosons.

\subsection{Split-SUSY loop effects in Higgs productions at the ILC}

\begin{figure}[hbt]
\epsfig{file=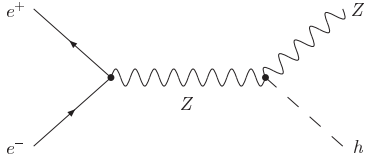,width=5cm} \vspace*{-0.4cm} \caption{Feynman
diagrams for $e^+e^-\to Z h$ at tree-level.} \label{eezh-tree}
\end{figure}

\begin{figure}[hbt]
\vspace*{-0.4cm} \epsfig{file=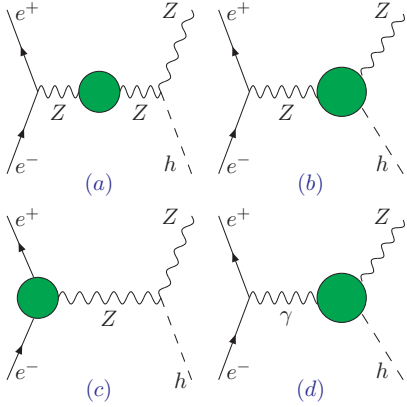,width=6cm} \vspace*{-0.4cm}
\caption{Feynman diagrams for $e^+e^-\to Z h$ with one-loop
corrected propagators and
         effective vertices in split-SUSY.}
\label{eezh-loop1}
\end{figure}

\begin{figure}[hbt]
\vspace*{-0.4cm} \epsfig{file=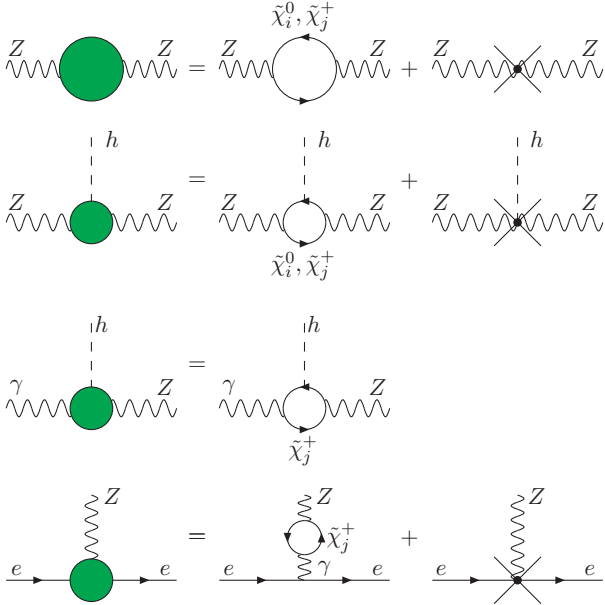,width=8.5cm}
\vspace*{-0.4cm} \caption{Feynman diagrams for each one-loop
corrected propagator and effective vertex in Fig.
\ref{eezh-loop1}.} \label{eezh-loop2}
\end{figure}

The tree-level  $e^+e^-\to Z h$ process is shown in Fig.
\ref{eezh-tree}. For the one-loop effects of split SUSY, we need
to calculate the diagrams containing the effective $Z$-boson
propagator and several effective vertices shown in Fig.
\ref{eezh-loop1}. Note that the box diagrams always involve
sfermions in the loops and thus drop out since all sfermions are
superheavy in split SUSY. In our calculations we use the on-shell
renormalization scheme \cite{denner}. For each effective vertex or
$Z$-boson propagator, we need to calculate several loops plus the
corresponding counterterms. For the new rare vertices induced at
loop level, such as $\gamma Zh$, there are no corresponding
counterterms. Since in split-SUSY all scalar superparticles are
superheavy and decouple from this process, the loops only involve
charginos and neutralinos, as shown in Fig. \ref{eezh-loop2}.

\begin{figure}[hbt]
\epsfig{file=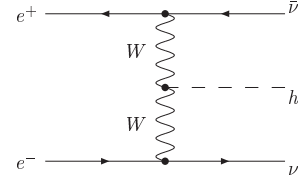,width=4cm} \vspace*{-0.3cm} \caption{Feynman
diagrams for $WW$-fusion process $e^+e^-\to h \nu_e \bar\nu_e $ at
tree-level.} \label{ww-tree}
\end{figure}

\begin{figure}[hbt]
\epsfig{file=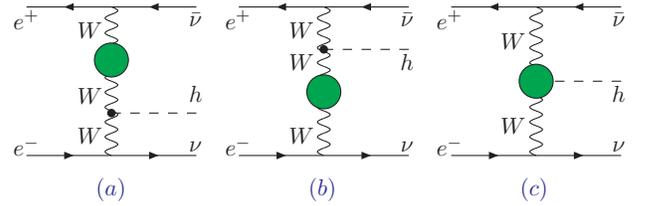,width=8.5cm} \vspace*{-0.5cm} \caption{Feynman
diagrams for $WW$-fusion process $e^+e^-\to \nu_e \bar\nu_e h$
with one-loop
         corrected propogators and effective vertices.}
\label{ww-loop1}
\end{figure}

\begin{figure}[hbt]
\epsfig{file=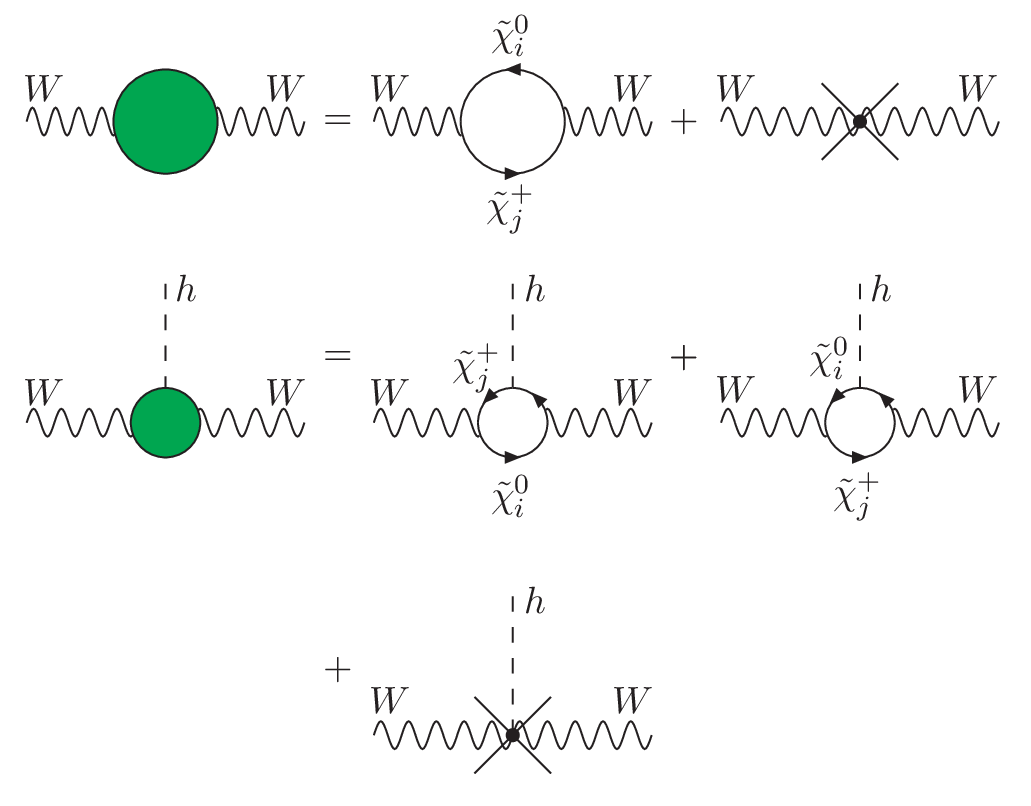,width=8cm} \vspace*{-0.5cm} \caption{Feynman
diagrams for each one-loop corrected propagator and effective
vertex in Fig. \ref{ww-loop1}.} \label{ww-loop2}
\end{figure}

For the $WW$-fusion process $e^+e^-\to \nu_e \bar\nu_e h$ our
calculations are similar as for $e^+e^-\to Z h$.  The tree-level
Feynman diagram is shown in Fig. \ref{ww-tree} and for one-loop
split-SUSY effects we need to calculate the diagrams containing
the effective $W$-boson propagator and several effective vertices
shown in Fig. \ref{ww-loop1}. Just like the diagrams shown in Fig.
\ref{eezh-loop2}, each effective vertex or $W$-boson propagator
contains several loops plus the corresponding counterterms, as
shown in Fig. \ref{ww-loop2}.

Note that for $e^+e^-\to \nu_e \bar\nu_e h$, in addition to the
$WW$-fusion contribution shown in Fig. \ref{ww-tree}, another
contribution comes from Higgs-strahlung process $e^+e^-\to Z h$
followed by $Z\to  \nu_e \bar\nu_e$. The cross section of
$e^+e^-\to Z h \to  \nu_e \bar\nu_e h$ peaks at the threshold of
$\sqrt s= M_Z+M_h$ and then falls rapidly as $\sqrt s$ increases,
where $\sqrt{s}$ is the center-of-mass (c.m.) energy of $e^+e^-$
collision. By contrast, the cross section of $WW$-fusion process
grows monotonously as $\sqrt s$ increases and is far dominant over
$e^+e^-\to Z h \to  \nu_e \bar\nu_e h$ for $\sqrt{s}\gg M_h$. In
our calculation we assume $\sqrt{s}=1$ TeV ($\gg M_h$) and thus we
only consider $WW$-fusion process.

Note that in the literature \cite{rewwh} the supersymmetric corrections to this
$WW$-fusion process have been computed, but those calculations focus on
the loops involving sfermions (squarks and sleptons).
In our calculations in the
scenario of split-SUSY, we consider the loops involving charginos and neutralinos,
ignoring the loops involving sfermions since all sfermions are superheavy in split-SUSY.
So far in the literature such chargino/neutralino loop corrections have not been
reported.

Each loop diagram
is composed of scalar loop functions \cite{Hooft} which are calculated by using
LoopTools \cite{Hahn}.
The calculations of the loop diagrams are tedious and the analytical expressions
are lengthy, which are not presented here.

\section{Numerical results}
In split-SUSY the masses of squarks and the CP-odd Higgs boson $A$
are assumed to be arbitrarily superheavy. As our previous study
showed \cite{wenyu}, their effects in low energy processes will
decouple as long as they are heavier than about 10 TeV. The Higgs
mass $M_h$ can be calculated from Feynhiggs \cite{feynhiggs} and
in our calculations we assume the masses of squarks and  Higgs
boson $A$ are 200 TeV. Among the low-energy parameters of
split-SUSY, i.e., $\tan\beta$, $M_2$, $M_1$ and $\mu$, $M_h$ is
sensitive to $\tan\beta$ and a large $\tan\beta$ leads to a large
$M_h$. In our calculations we fix  $\tan\beta=40$ since a large
value of  $\tan\beta$ is favored by current experiments. Our
results are not sensitive to $\tan\beta$ in the region of large
$\tan\beta$ value and our results are approximately valid for
$\tan\beta \gtrsim 10$. With the input values of  $\tan\beta$ and
squark masses, we get $M_h=120$ GeV from Feynhiggs
\cite{feynhiggs}.

With the fixed value of $\tan\beta$, there remained three split-SUSY parameters:
$M_2$, $M_1$ and $\mu$.  We further use the unification relation
$M_1=5 M_2\tan^2\theta_W/3 \simeq 0.5 M_2$,
which is predicted in the minimal supergravity model. Thus finally we have two
free SUSY parameters. The SM parameters used in our results are taken from \cite{pdg}.

\subsection{Numerical results without WMAP constraints}
\begin{figure}[hbt]
\epsfig{file=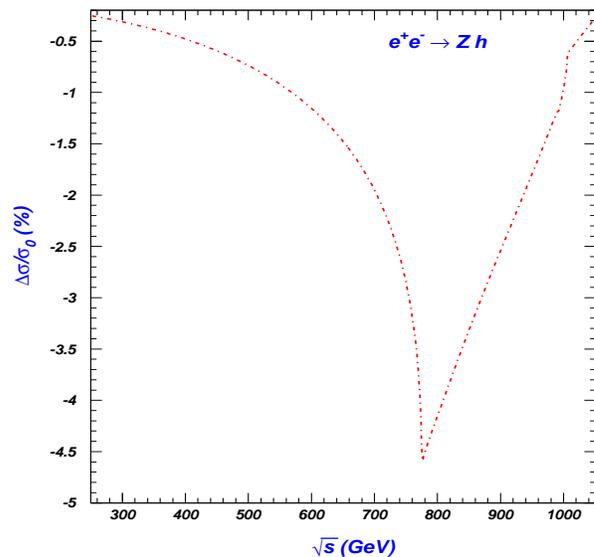,width=8cm,height=7.5cm} \vspace*{-0.5cm}
\caption{The relative one-loop correction of split-SUSY to the
cross section of
         $e^+e^-\to Z h$ versus the c.m. energy. }
\label{fig7}
\end{figure}

In order to show the features of our results, we first present
some results without considering the WMAP dark matter constraints.
In Fig. \ref{fig7} we show the relative one-loop correction of
split-SUSY to the cross section of $e^+e^-\to Z h$ versus the c.m.
energy of $e^+e^-$ collision for $M_2=400$ GeV and $\mu=600$ GeV.
In this case the lightest chargino mass $M_{\tilde \chi^+_1}=387$
GeV. We see from  Fig. \ref{fig7} that the corrections are
negative and have a peak at $\sqrt s= 2 M_{\tilde \chi^+_1}$ due
to the threshold effects. The magnitude of the corrections for
$\sqrt s= 1$ TeV, which will be taken for our following studies,
is relatively small.

\begin{figure}[hbt]
\vspace*{-0.5cm}
\epsfig{file=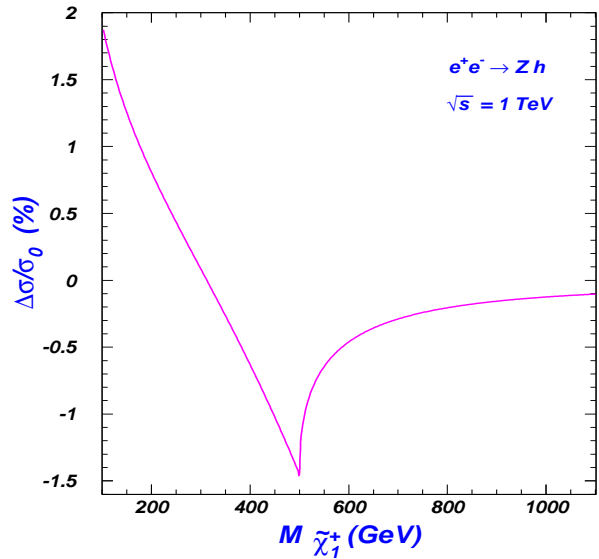,width=8cm,height=7.5cm}
\vspace*{-0.5cm}
\caption{Same as Fig.\ref{fig7}, but versus the chargino mass for
         the c.m. energy of 1 TeV.}
\label{fig8}
\end{figure}

In Fig. \ref{fig8} we fix $\sqrt s= 1$ TeV and  $\mu=100$ TeV
(note that the scenario with a very large $\mu$ is proposed and
argued in \cite{split-split}), and by varying $M_2$  we show the
relative one-loop correction of split-SUSY to the cross section of
$e^+e^-\to Z h$ versus  the lightest chargino mass $M_{\tilde
\chi^+_1}$ (in this case the  chargino mass $M_{\tilde \chi^+_1}$
is almost equal to $M_2$ due to the superheavy higgsinos). The
peak happens at $M_{\tilde \chi^+_1}=\sqrt s/2$ due to threshold
effects. When the chargino mass gets heavier than 1 TeV, the
corrections becomes very small, showing the decoupling property.

\subsection{Numerical results with WMAP constraints}
Now we require the lightest neutralinos make up the cosmic dark matter relic
density measured by WMAP, which is given by $0.085<\Omega_{CDM}h^2<0.119$ at
$2\sigma$ \cite{wmap} with $h=0.73$ being the Hubble constant.
Of course, the direct bounds from LEP experiments \cite{lep2-web} need to
be also considered, which are:
(i) the lightest chargino heavier than about 103 GeV;
(ii) the lightest neutralino heavier than about 47 GeV;
(iii) $\tan\beta$ larger than 2.
Note that the LEP bound $\tan\beta>2$ is obtained from the search limit of
the lightest Higgs boson for squarks below 1 TeV. Such a bound may be
relaxed in split-SUSY because of superheavy squarks.

We then perform a scan over the parameter space of $M_2$ and $\mu$.
The $2\sigma$ allowed region is shown in Fig. 2 of Ref. \cite{wenyu}.
(Note that in \cite{wenyu} we used the one-year WMAP data
$0.094 < \Omega_{CDM}h^2 < 0.129$. The allowed region
with one-year WMAP data is approximately same as that with
three-year WMAP data).

\begin{figure}[hbt]
\epsfig{file=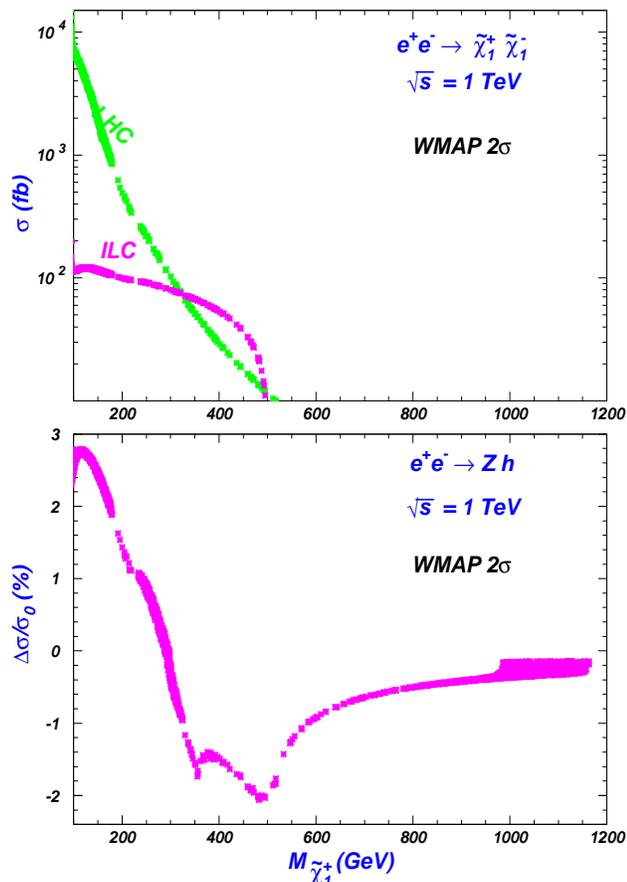,width=8.5cm,height=12cm}
\vspace*{-0.5cm}
\caption{The shaded areas are the 2$\sigma$ region of split-SUSY parameter space allowed by
         the WMAP
         dark matter measurement in the planes of the chargino pair production
         rate (upper panel) and  the one-loop correction of
         split-SUSY to the cross section of $e^+e^-\to Z h$ (lower panel)
         versus the chargino mass.}
\label{fig9}
\end{figure}

In Fig. \ref{fig9} we show the one-loop correction of split-SUSY
to the cross section of $e^+e^-\to Z h$ (lower panel) with
comparison to the chargino pair production rate (upper panel). The
chargino pair production rate is calculated at tree-level, as in
our previous work  \cite{wenyu}.

From Fig. \ref{fig9} we see that when the chargino is lighter than about 300 GeV, the
chargino pair production rate at the ILC is large and the corresponding virtual effects
in $e^+e^-\to Z h$ are positive. When the chargino gets heavier,  the  chargino pair
production rate at the ILC drops rapidly. Of course, when the chargino is heavier than
500 GeV, beyond the threshold of the ILC (with c.m. energy of 1 TeV), the charginos
cannot be pair produced. Then it is interesting to observe that for a chargino between
500 and 600 GeV, although the ILC cannot produce chargino pairs, the virtual effects in
$e^+e^-\to Z h$ can still reach a couple of percent in magnitude and thus may be observable
at the ILC with a high integrated luminosity. Finally, when the chargino is heavier than
about 600 GeV,  it will probably remain unaccessible because both the chargino pair production
rates and the virtual effects are very small due to the decoupling property of SUSY.

Note that for $e^+e^-\to Z h$
we numerically compared our results with the full one-loop corrections given in
\cite{complete-susy} (we thank the authors of \cite{complete-susy} for giving
us their fortran code). In our calculations we only considered the chargino and
neutralino loops, while in their calculations the sfermion loops are also
considered besides the chargino and neutralino loops. In principle, their results
in the limit of superheavy sfermions should approach to our results. We found that
although their fortran code does not work well for superheavy sfermions (say above 10 TeV)
due to the limitation of numerical calculation, for a given point in
the parameter space our results agree well with those by using
their fortran code with all sfermions above 1 TeV.

\begin{figure}[hbt]
\epsfig{file=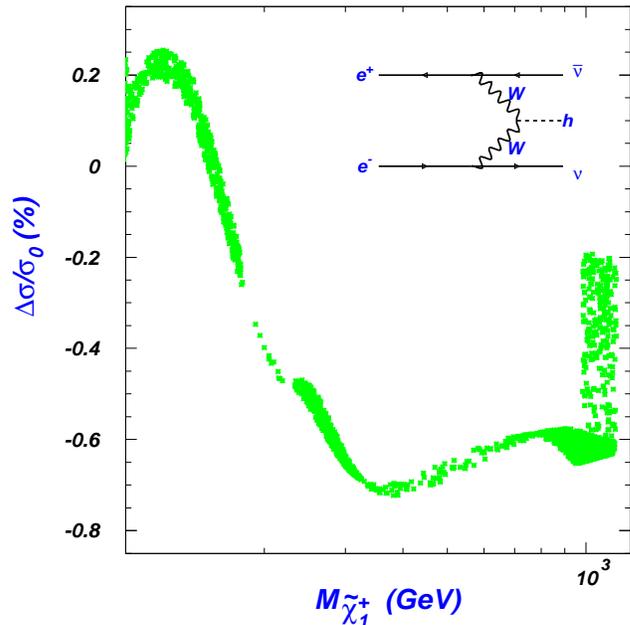,width=8.5cm,height=8.5cm}
\vspace*{-0.5cm}
\caption{Same as the lower panel of Fig.\ref{fig9}, but for the $WW$-fusion process
         $e^+e^-\to \nu_e \bar\nu_e h$.}
\label{fig10}
\end{figure}

The  one-loop correction of split-SUSY to the cross section of
$WW$-fusion process $e^+e^-\to \nu_e \bar\nu_e h$ is very small in
magnitude, below one percent, as shown in  Fig. \ref{fig10}. Even
with a high luminosity the ILC can hardly reveal such a small
deviation from the measurement of this process. The reason why the
virtual effects in the $s$-channel process $e^+e^-\to Z h$ is much
larger in magnitude than in  the $t$-channel process $e^+e^-\to
\nu_e \bar\nu_e h$ may be that for the $s$-channel process the
virtual sparticles (charginos and neutralinos) in the loops could
be more energetic and cause larger quantum effects.

Anyway, such virtual effects of split-SUSY, no matter large or
small in magnitude, could be informative and complementary to the
real sparticle productions in probing split-SUSY at colliders. For
example, if split-SUSY turns out to be the true story and the
chargino pair production is observed with the chargino mass around
150 GeV at the ILC, then we know from  Figs. \ref{fig9} and
\ref{fig10} that the virtual effects of SUSY must be about $2.5\%$
for  process $e^+e^-\to Z h$ and $-0.1\%$ for $WW$-fusion process
$e^+e^-\to \nu_e \bar\nu_e h$.

\section{Conclusion}
In split supersymmetry, gauginos and higgsinos are the only supersymmetric
particles possibly accessible at foreseeable colliders like the
LHC and the ILC.
In order to account for the cosmic dark matter measuerd by WMAP, the parameter space
of the gauginos and higgsinos in split supersymmetry are stringently constrained,
which can be explored at the LHC and the ILC through direct productions and the
virtual effects of these gauginos and higgsinos.
The clean environment of the ILC may render the virtual effects of percent level
meaningful in probing the new physics.
In this work we assumed split supersymmetry and calculated the virtual effects
of the WMAP-allowed gauginos and higgsinos in Higgs productions $e^+e^-\to Z h$ and
$e^+e^-\to \nu_e \bar\nu_e h$  through $WW$ fusion at the ILC.
We found that the production cross section of $e^+e^-\to Zh$ can be altered by a
few percent in some part of the WMAP-allowed parameter space, while
the correction to the $WW$ fusion process $e^+e^-\to \nu_e \bar\nu_e h$
is below $1\%$.

Such virtual effects are correlated with
the cross sections of chargino pair productions and thus can offer complementary
information in probing split supersymmetry at the colliders. Our results indicate
that if the lightest chargino is in the light region allowed by the WMAP dark matter
(say below 200 GeV), then at the ILC and LHC the chargino pair production rates
are large and the virtual effects of charginos/neutralinos in the process $e^+e^-\to Z h$
at the ILC can reach a few percent, both of which may be measurable and cross-checked.
An interesting observation is that for a chargino between
500 and 600 GeV, although the ILC  (with c.m. energy of 1 TeV) cannot produce chargino pairs,
the virtual effects in
$e^+e^-\to Z h$ can still reach a couple of percent in magnitude and thus may be observable
at the ILC with a high integrated luminosity.
The WMAP-allowed region with the chargino heavier than about 600 GeV
will most likely remain unaccessible because both the chargino production rates
and the virtual effects are very small due to the decoupling property of SUSY.

\vspace*{0.5cm}

This work is supported in part by National Natural Science
Foundation of China.


\begin{thebibliography}{11}
\bibitem{ILC} K. Abe {\it et al.}, hep-ph/0109166;
              T. Abe {\it et al.}, hep-ex/0106056;
              J. A. Aguilar-Saavedra, {\it et al.}, hep-ph/0106315.
\bibitem{split} N. Arkani-Hamed, S. Dimopoulos, hep-th/0405159;
                G.F. Giudice, A. Romanino, \NPB699, 65 (2004);
                N. Arkani-Hamed, S. Dimopoulos, G. F. Giudice, A. Romanino, \NPB709, 3 (2005).
\bibitem{senatore} L. Senatore, \PRD71, 103510  (2005).
\bibitem{wang} F. Wang, W. Y. Wang, J. M. Yang, \PRD72, 077701 (2005).
\bibitem{wenyu} F. Wang, W. Y. Wang, J. M. Yang, \EPJC46, 521 (2006).
\bibitem{pierce} A. Pierce, \PRD70,  075006 (2004);
                 A. Arvanitaki, P. W. Graham, hep-ph/0411376.
\bibitem{profumo}  A. Masiero, S. Profumo, P. Ullio, \NPB712, 86 (2005).
\bibitem{gluino-lhc} For the studies of split-SUSY gluino at LHC, see, e.g.,
                    K. Cheung, W.-Y. Keung, \PRD71, 015015 (2005);
                    J. G. Gonzalez, S. Reucroft, J. Swain, \PRD74,  027701 (2006).
\bibitem{higgs-residue} SUSY-QCD may have large residue effects in Higgs processes,
                        see, e.g.,
                        G. Gao, R. J. Oakes, J. M. Yang, \PRD71, 095005 (2005);
                        J. Cao,  {\it et al.}, \PRD68, 075012 (2003);
                        G. Gao,  {\it et al.},  \PRD66, 015007 (2002).
                       H.~E.~Haber,  {\it et al.}, \PRD63, 055004 (2001);
                       M.~J.~Herrero, S.~Pe\~naranda and D.~Temes, \PRD64, 115003 (2001);
                       A. Dobado, M.~J.~Herrero, \PRD65, 075023(2002);
                       M.~Carena,  {\it et al.}, \PRD60, 075010 (1999); \PRD62, 055008 (2000).
\bibitem{top-susy} For SUSY-QCD effects in $t\bar t$ productions,
                   see, e. g.,
                    C. S. Li,  {\it et al.}, \PRD52, 5014 (1995);  \PLB379, 135 (1996);
                    S. Alam, K. Hagiwara, S. Matsumoto, \PRD55, 1307 (1997);
                    Z. Sullivan, \PRD56, 451 (1997);
                    For SUSY-QCD effects in FCNC top interactions,
                    see, e. g.,
            C.~S.~Li, R.~J.~Oakes, J.~M.~Yang, \PRD49, 293 (1994);
                    G.~Couture, C.~Hamzaoui and H.~Konig, \PRD52, 1713 (1995);
                    J.~L.~Lopez, D.~V.~Nanopoulos and R.~Rangarajan, \PRD56, 3100  (1997);
                    G.~M.~de Divitiis, R.~Petronzio and L.~Silvestrini, \NPB504, 45 (1997);
                     C.~S.~Li, {\it et al.}, \PLB599, 92 (2004);
                    J. Cao, {\it et al.}, \NPB651, 87 (2003); \PRD74, 031701 (2006).
                    M. Frank, I. Turan, \PRD74,  073014 (2006).
\bibitem{complete-susy} P.Chankowski, S.Pokorski, J.Rosiek, \NPB423, 437  (1994).
\bibitem{rewwh} T. Hahn et al., \NPB652, 229 (2003);
                H. Eberl et al., \NPB657,378 (2003).
\bibitem{gunion}  H. E. Haber and G. L. Kane, Phys. Rep. {\bf 117}, 75 (1985);
                  J. F. Gunion and H. E. Haber, \NPB272, 1 (1986).
\bibitem{higgsmass}  H.E. Haber, R. Hempfling, \PRL66, 1815 (1991);
                  Y. Okada, M. Yamaguchi, T. Yanagida, Prog. Theor. Phys. {\bf 85}, 1 (1991);
                                                                            \PLB262,54(1991);
                 J. Ellis, G. Ridolfi, F. Zwirner, \PLB257, 83 (1991); \PLB262, 477 (1991);
                 J.~R.~Espinosa and R.~J.~Zhang, JHEP {\bf 0003} (2000) 026.
\bibitem{denner} A. Denner, Fortschr. Phys. {\bf41} (1993)4
\bibitem{Hooft} G.~'t Hooft and M.~J.~G.~Veltman, \NPB153, 365 (1979).
\bibitem{Hahn} T.~Hahn and M.~Perez-Victoria,  Comput.\ Phys.\ Commun.\  {\bf 118}, 153 (1999);
               T.~Hahn,  Nucl.\ Phys.\ Proc.\ Suppl.\  {\bf 135}, 333 (2004).
\bibitem{feynhiggs} S.Heinemeyer, hep-ph/0407244
\bibitem{pdg} W.~M.~Yao {\it et al.}, \JPG33, 1 (2006).
\bibitem{split-split} K. Cheung, C.-W. Chiang, \PRD71, 095003 (2005).
\bibitem{wmap}  D. N. Spergel, {\it et al.}, astro-ph/0603449.
\bibitem{lep2-web} LEP2 SUSY Working Group homepage, http://lepsusy. web.cern.ch/lepsusy/
\end{thebibliography}
\end{document}